# Thinging the Use Case Model

Sabah Al-Fedaghi
Computer Engineering Department
Kuwait University
Kuwait
sabah.alfedaghi@ku.edu.kw

*Abstract*—Use cases as textual visual modeling techniques have become a key construct and the foundation of the most popular de facto standard technique for performing software requirements analysis and specification. This paper describes use cases in terms of a recently proposed model: the thinging machine (TM) model. Such a link to TM strengthens the notion of use cases and clarifies related concepts. For example, the concept of events is utilized in an unconstrained way in use cases, and TM may contribute in this direction. Several selected use cases are remodeled using TM. This study provides many insights. For example, one result shows that use cases are an activation apparatus of "mega-triggering" (high-level events) wherein a group of submachines (processes) are actuated.

*Keywords-conceptual modeling; use case; thinging machine; generic processes, events*

## I. INTRODUCTION

Since their initial introduction by Jacobson [1] in 1987, use cases as textual visual modeling techniques have become a key construct and the foundation of the most popular de facto standard technique for performing software requirements analysis and specifications [2]. A use case is a description of a set of possible interactions that an individual actor initiates with a system. An actor is a role played by a user [3]. The description of a use case can typically be divided into a basic course and zero or more alternative courses. It "describes a sequence of actions, performed by a system, that yields a result of value to a user" [2]. Use case modeling is a technique that is quite independent from object modeling. According to [4], "Use-case modeling can be applied to any methodology or object-oriented. It is a discipline of its own, orthogonal to object modeling."

In UML, the use case diagram is the primary form of system/software requirements [5], and other UML diagram types are linked from use cases. A use case is an effective technique for communicating system behavior by specifying system behavior: "One of the beauties of use cases is their accessible, informal format. Use cases are easy to write, and the graphical notation is trivial" [6].

Use case modeling has its well-known problems [7] [8-9] [10]. The use cases' simplicity can be deceptive; many groups writing use cases for the first time run into problems [6]. Putting the use case approach into practice also often illuminates problems that are not addressed in books and most articles on use cases, and different requirements engineers typically perform use case modeling differently [10] [3]. According to Leffingwell and Widrig [3], "Use cases should be only one of several ways of capturing user requirements."

Use cases have been widely discussed for over thirty years, and there is no need to further discuss them here. The purpose of this paper is to posit that the recently proposed thinging machine (TM) modeling methodology can be utilized as a tool to improve the notion of use cases. Until we introduce TM in the next section, the following example illustrates our general outline for using TM in the context of use cases.

### A. Example of TM Model for Use Cases

According to Visual Paradigm [5], "A Use Case diagram illustrates a set of use cases for a system, i.e. the actors and the relationships between the actors and use cases." Fig. 1 presents a sample use case.

The TM models the totality of the involved portion of reality using only the following notions:
1. Five generic processes: Create, Process, Release, Transfer, and Receive.
2. Solid arrows to represent the flow of things.
3. Dashed arrows to represent triggering.

Fig. 2 shows the TM model that corresponds to the use case. The student creates a request for a book (circle 1–assuming all necessary information is included in the request).

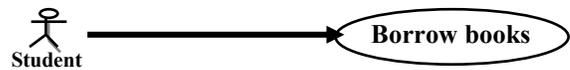

**Figure 1. A simple use case (*redrawn from [5]*).**

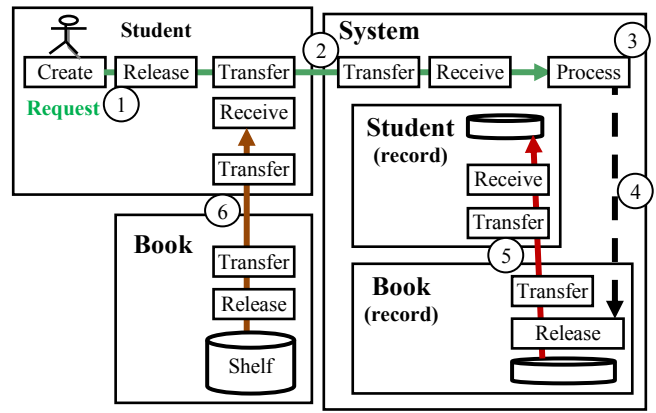

**Figure 2. The TM model of the use case.**

The request (circle 1 in the figure) flows (2) to the system, where it is then processed (3) to trigger (4) adding the data about the book to the student record about borrowed books (5). This is supposed to mirror an actual process in reality in which a book flows to the hands of the student (6). Fig. 3 shows a simplification of the TM model, where we assume that the direction of the arrow indicates the direction of the flow; thus, we can eliminate Release and Transfer.

Fig. 4 shows a further simplification by eliminating Create, Process, and Receive. The start of an arrow indicates Create, the start of triggering indicates Process, and the end of an arrow indicates Receive. Fig. 5 shows a reduction of the TM diagram to the use case diagram by eliminating details and some structures.

This paper connects use cases to the TM model. We claim that such a link strengthens the notion of use cases and clarifies related concepts.

*B. Some Justification to Apply TM to Use Cases*

Consider the classical book *Managing Software Requirements: A Use Case Approach* [2], in which the term *event* is used numerous times. The following quotes are from that book (italics added).

> The use cases—sequences of *events* by which the actors interact with the business elements to get their job done.

> The use case specifications can be thought of as a container that describes a series of related *events*.

> These use cases are so simple that they do not need a detailed description of the flow of *events*.

> The heart of the use case is the *event flow*, usually *a textual description of the operations* by the actor and the system's various responses. Therefore, the *event flow* describes what the system is supposed to do based on the actor's behavior. By the way, it is not required that the flow be described textually. You can use UML interaction diagrams.

> The *flow of events* conveys the meat of the use case's purpose.

> *Flow of events* does not specify how the system does any of those things. It specifies only what happens.

> A use case that prints a receipt for a credit card transaction may discover that the printer has run out of paper. This special case would be described within the use case as an alternative *flow of events*.

> It is important to distinguish between *events* that start the use case flows and preconditions.

> In Finite State Machines, both the output and the next state can be determined solely on the basis of understanding the current state and the *event* that caused the transition.

> [Sample of *events*] On press, Count press, Bulb burns out, Every second.

> Although use cases show how requirements can drive design, it is also true that we can model the implementation of any individual requirement, or any set of requirements, as collaboration.

> Although the use case does have some special properties—namely, the *sequence of events*—we can often arrange our itemized requirements to accomplish the same objective.

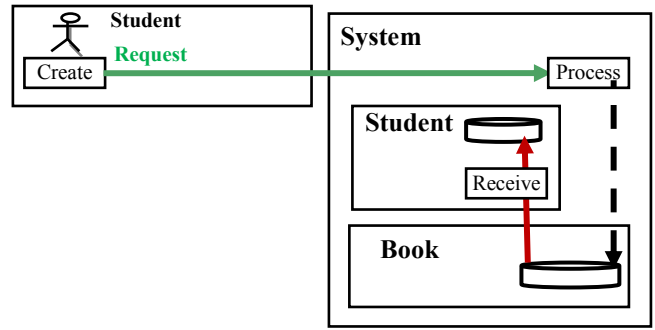

Figure 3. Simplification of the TM model.

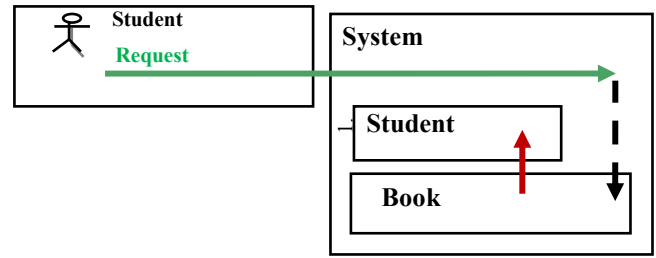

Figure 4. Further simplification of the TM model.

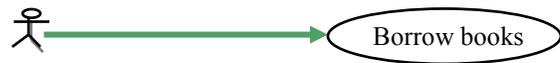

Figure 5. Use case diagram that resulted from TM model simplification.

> Interactions with the system closely mimic the *real-world flow of events*.

> An *event-time-schedule* shall be 1 minute.

> Basic *flow of events*.

> *Event*s associated with the alternative behavior.

> Complex processes [such as] sequences of activities, state transitions, *events*.

Accordingly,
- Events seem to be *complex processes* associated with *behavior* and *time*, and they seem to cause *transitions*.
- Events are special *properties* of use cases.
- Flow of events specifies only *what happens*.

The book *Managing Software Requirements: A Use Case Approach* [3] uses *flow* of events and such terms as a use case *flow*, cash *flow*, work *flow*, basic *flow*, or thread, *flow*charts, data *flow*, and process *flow*.

It seems that the notion of an event is utilized in an unconstrained use way, and TM may assist in this direction. TM defines an *event* as a process (called a machine) that includes the following:

- The region of the event,
- The time of the event,
- The event itself, and
- Other things that will not be discussed here.

For example, we modeled the event *A student requests a book* in the previously discussed example, as shown in Fig. 6. For simplicity's sake, we will represent an event only by its region. Accordingly, we can select the following events for the book-borrowing example:

Event 1 ($E_1$): *A student requests a book.*
Event 2 ($E_2$): *The request is processed.*
Event 3 ($E_3$): *The borrowing is recorded in the system.*
Event 4 ($E_4$): *The student receives the requested book.*

Fig. 7 shows these events. Then, we specify the behavior of this use case, as shown in Fig. 8.

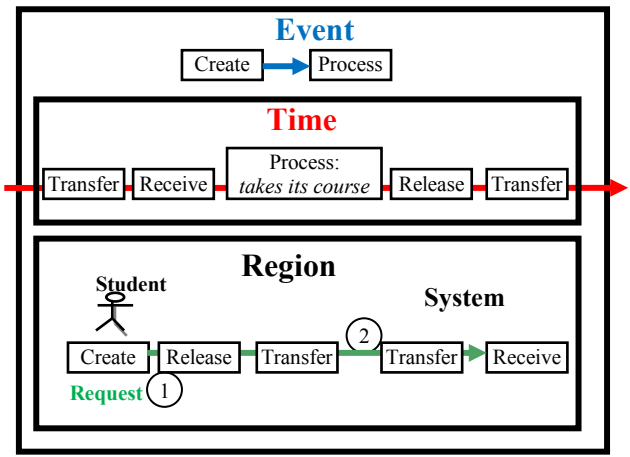

**Figure 6. The event *A student has requested a book*.**

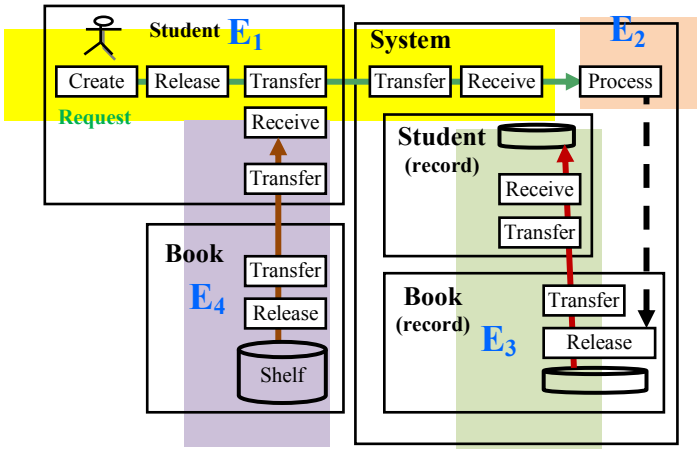

**Figure 7. The TM model of the use case.**

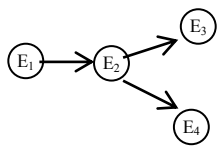

**Figure 8. The behavior of the book-borrowing use case.**

This paper uses TM modeling for use cases, as described above. We claim that such a venture will benefit requirement modeling in software engineering, especially in the context of use cases.

In the next section, we will review the TM to achieve a self-contained paper. The rest of the paper involves TM remodeling of some use cases selected from the literature.

## II. THINGING MACHINE

The TM model has been applied to many real systems such as phone communication [11], physical security [12], vehicle tracking [13], intelligent monitoring [14], asset management [15-16], information leakage [17], engineering plants [23], inventory management processes [18], procurement processes [19], public key infrastructure network architecture [20], bank check processing [21], wastewater treatment [22], computational thinking [23], computer attacks [24], provenance [25], services in banking industry architecture networks [26], digital circuits [27], chip manufacturing [28], service-oriented systems [29] and programming [30], and unmanned aerial vehicles [31].

The notion of *thinging* has its origin in Heidegger's work on *The Thing* [32]. A TM is an abstract machine that has five generic stages: Create, Process, Release, Transfer, and Receive things, as shown in Fig. 9. Fig. 10 shows a TM in the classical input→process→output form. Details of TMs can be found in the references [32-35].

The generic operations on things and by machines can be described as follows:

**Arrive**: A thing reaches a new machine.
**Accepted**: A thing is permitted to enter the machine.
If arriving things are always accepted, Arrive and Accept can be combined as a **Received** stage. For the purpose of simplification, the examples in this paper assume the received stage exists without loss of generality.
**Processed** (changed): A thing undergoes some kind of transformation that changes it without creating a new thing.
**Released**: A thing is marked as ready to be transferred outside the machine.

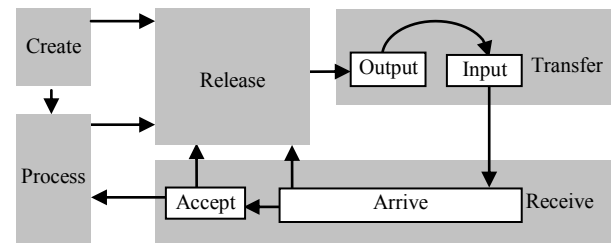

**Figure 9. TM.**

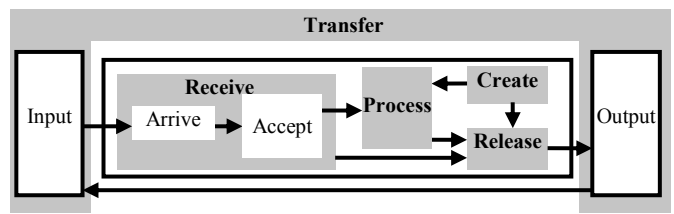

**Figure 10. Another form of a TM.**

**Transferred**: A thing is transported somewhere to/from outside the machine.

**Created**: A new thing is born (created) in a machine. The term Create comes from creativity with respect to a system (e.g., constructed things from already created things, or emergent things appearing from somewhere).

Flow (solid arrow) among stages is a spatiotemporal event, which signifies a thing's *conceptual* movement from one machine to another or among stages of a machine. Additionally, the TM model includes memory and triggering relations (represented as dashed arrows) among the processes' stages (machines).

### III. WHAT DOES A USE CASE LOOK LIKE?

Under the title *What does a use case look like?* Cockburnm [8] provides a use case entitled *Register arrival of a box*. In this use case, RA means "Receiving Agent," RO means "Registration Operator," and DS means "Department Supervisor." It is specified as follows:

Primary Actor: RA System:
Nighttime Receiving Registry Software
1. RA receives and opens box (box ID, bags with bag IDs) from TransportCompany (TC).
2. RA validates box ID with TC-registered IDs.
3. RA perhaps signs paper form for delivery person.
4. RA registers arrival into system, which stores the following: RA ID date, time box ID TC # bags (with bag IDs)
5. RA removes bags from box, places them onto the cart, and takes them to the RO. [8]

Fig. 11 shows the TM model for this case.
1. At nighttime (1), a delivery (2) from TC to RA is triggered.
2. A box (3) flows to RA (4), where it is opened (5) to trigger the following:
   - The box ID (6) and the TC-registered ID (7 and 8) are validated (9).

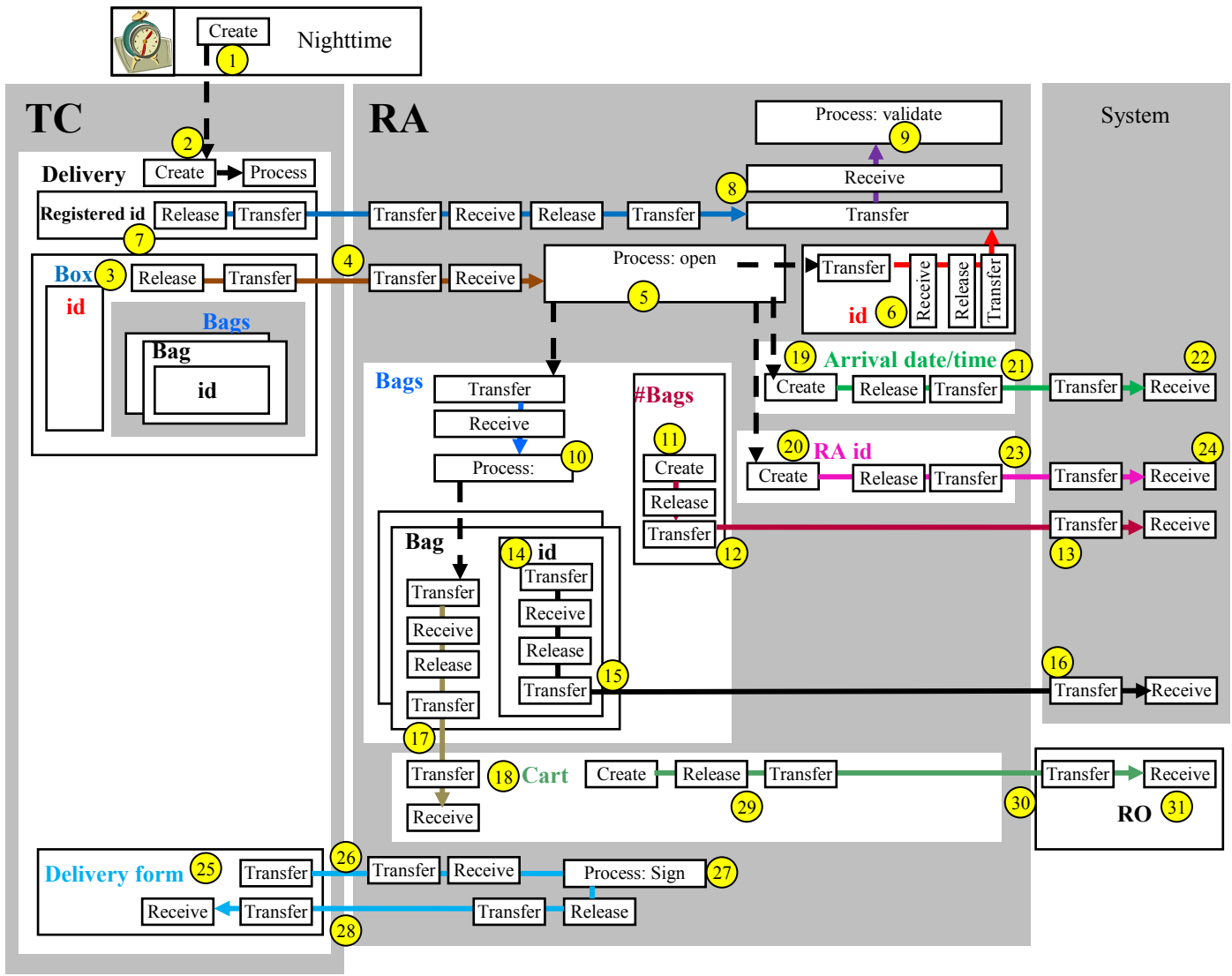

Figure 11. Use case register arrival of a box for the use case *Register arrival of a box*.

- The bags inside the box are processed, (10) and
  - The number of bags (11) is registered in the system (12 and 13).
  - For each bag,
    (i) The bag ID (14) is registered in the system (15 and 16).
    (ii) The bag is placed on (17) the cart (18).
- The arrival time (19) of the box and the RA ID (20) are registered in the system (21 and 22) and (23 and 24), respectively.
3. The delivery form (25) flows from TC to the RA (26) to be signed by the RA (27) and returned to TC (28).
4. The cart is driven (29) to the RO (30 and 31).

Accordingly, the events of the use case can be identified as follows (see Fig. 12):
Event 1 ($E_1$): At nighttime, a delivery from TC to RA occurs.
Event 2 ($E_2$): A box flows to the RA.
Event 3 (E3): The box ID and the TC-registered ID are validated.
Event 4 ($E_4$): The bags inside the box are processed.
Event 5 ($E_5$): The number of bags is registered in the system.
Event 6 ($E_6$): For each bag, the bag ID is registered in the system.
Event 7 ($E_7$): Each bag is placed onto the cart.
Event 8 ($E_8$): The arrival time of the box and the RA ID are registered in the system.
Event 9 ($E_9$): The delivery form flows from TC to the RA (26) to be signed by the RA and returned to TC.
Event 10 ($E_{10}$): The cart is driven to the RO.

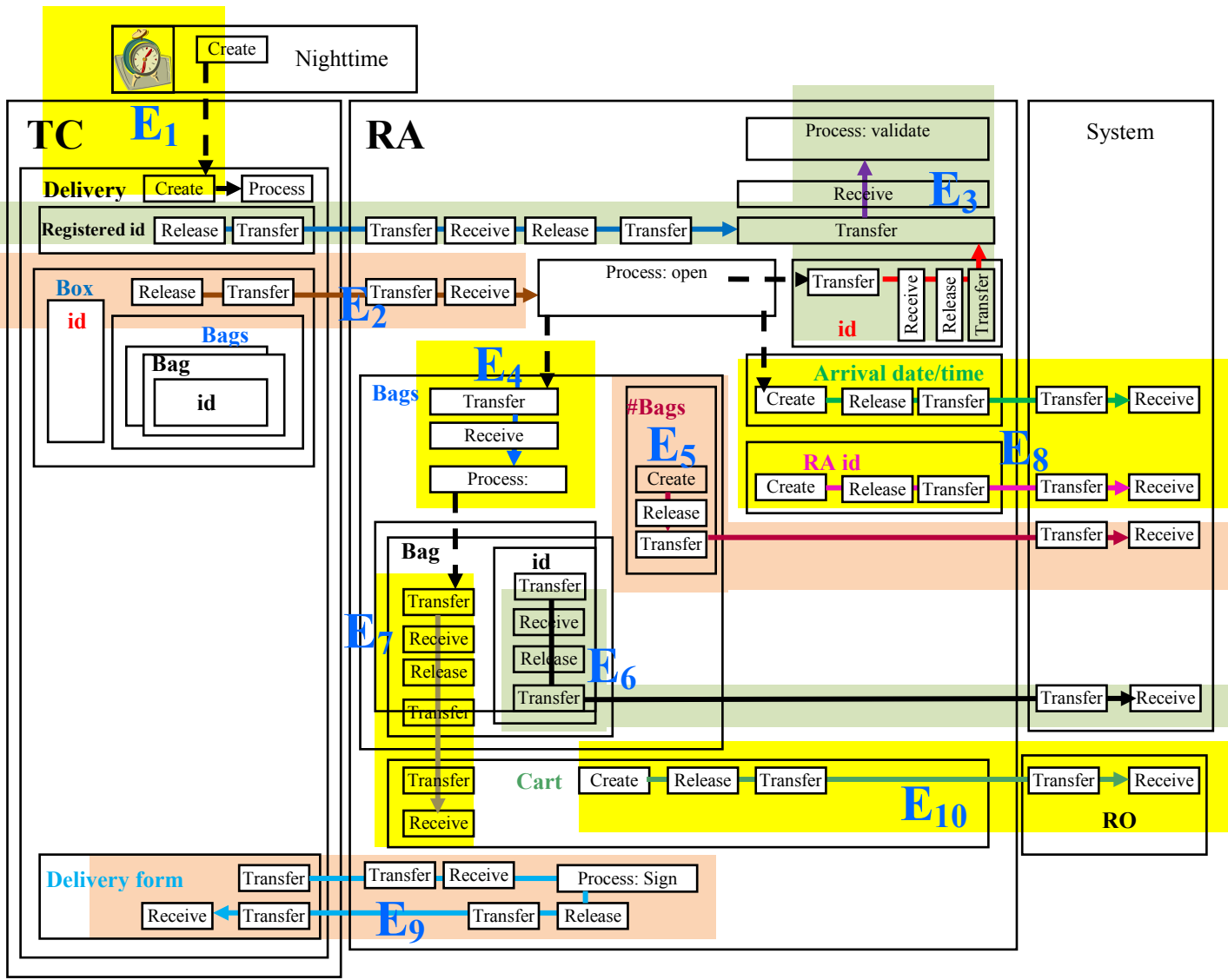

Figure 12. TM model of the use case *Register arrival of a box*.

Fig. 13 shows the behavior of the use case in terms of the chronology of events, which is called the flow of events in the use case literature (discussed in the introduction).

Various levels of simplification can be performed on the TM diagram. For example, in Fig. 14, Release, Transfer, and Receive are removed, assuming that the directions of arrows indicate the flow.

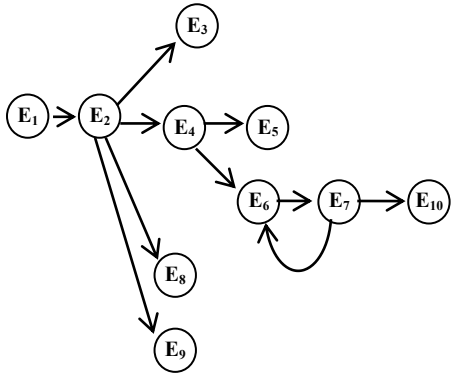

**Figure 13. The behavior of the use case *Register arrival of a box*.**

## IV. BASIC AND ALTERNATIVE USE CASES

Leffingwell and Widrig [2], in their book *Managing Software Requirements: A Use Case Approach*, present a step-by-step procedure for defining a use case. The use case presented involves a simple lighting control system called HOLIS. A resident activates a light in a house using the HOLIS. The authors provide a use case that prescribes the way in which lights are turned on or off or are dimmed based on how long the user presses a light switch in various manners.

The basic *Flow* for the Control Light use case is given as follows:

- Basic flow begins when the resident presses any button on the Control Switch. If the resident removes pressure from the Control Switch within the timer period, the system toggles the state of the light.
- If the light was on, the light is turned off, and there is no illumination.
- If the light was off, the light is turned on to the last remembered brightness level.

End of basic flow [2].

An alternative flow of events will occur when the resident holds a button on the Control Switch down for more than 1 second. Therefore, we need to add an alternative flow to the use case. If the resident keeps pressure on the Control Switch for more than 1 second, the system initiates a dimming activity for the indicated Control Switch button [2].

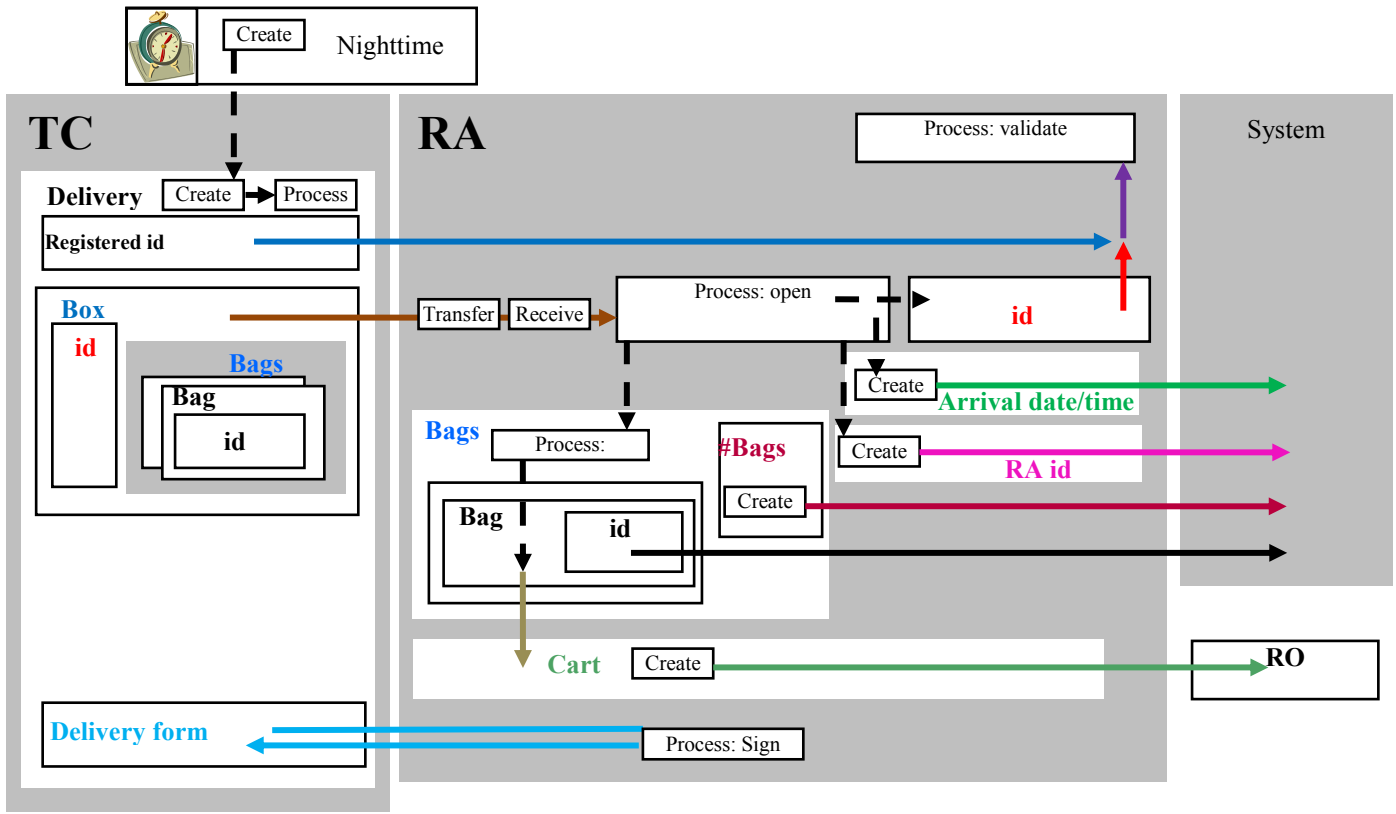

**Figure 14. Simplification of the use case *Register arrival of a box*.**

While the resident continues to press the Control Switch button,

1. The brightness of the controlled light is smoothly increased to a system-wide maximum value at a rate of 10% per second.
2. When the maximum value is reached, the brightness of the controlled light is smoothly decreased to a system-wide minimum value at a rate of 10% per second.
3. When the minimum value is reached, processing continues at subflow step 1.

When the resident stops pressing the Control Switch button, the system ceases to change the brightness of the light [2].

Fig. 15 shows the TM model of the use case. Note that, in a TM, a use case is a type of "mega-triggering" (high-level event) wherein a group of submachines are activated, as will be illustrated in this example. Fig. 15 includes the basic and alternative use cases, and each activates submachines (processes) included in the thick purple boxes that represent the region of the use case.

Upon the arrival (red star in the figure) of the analog signal (1), it is processed to convert it to a digital signal (2).

**The basic use case** is described as follows:
1. The digital signal is processed (3) to trigger processing of the light state (4).
   (i) If the current state is ON (5), then an OFF state is created (6).
   (ii) If the current is OFF (7), then
   (iii) The brightness level (8) is retrieved (9) and processed (10) to create an ON state (11).

**An alternative use case** is described as follows**.**
Under the two conditions,
(i) The light is on (11), and
(ii) The analog signal is still input (12; i.e., the resident is still pressing the button),
the alternative use case is activated (13).

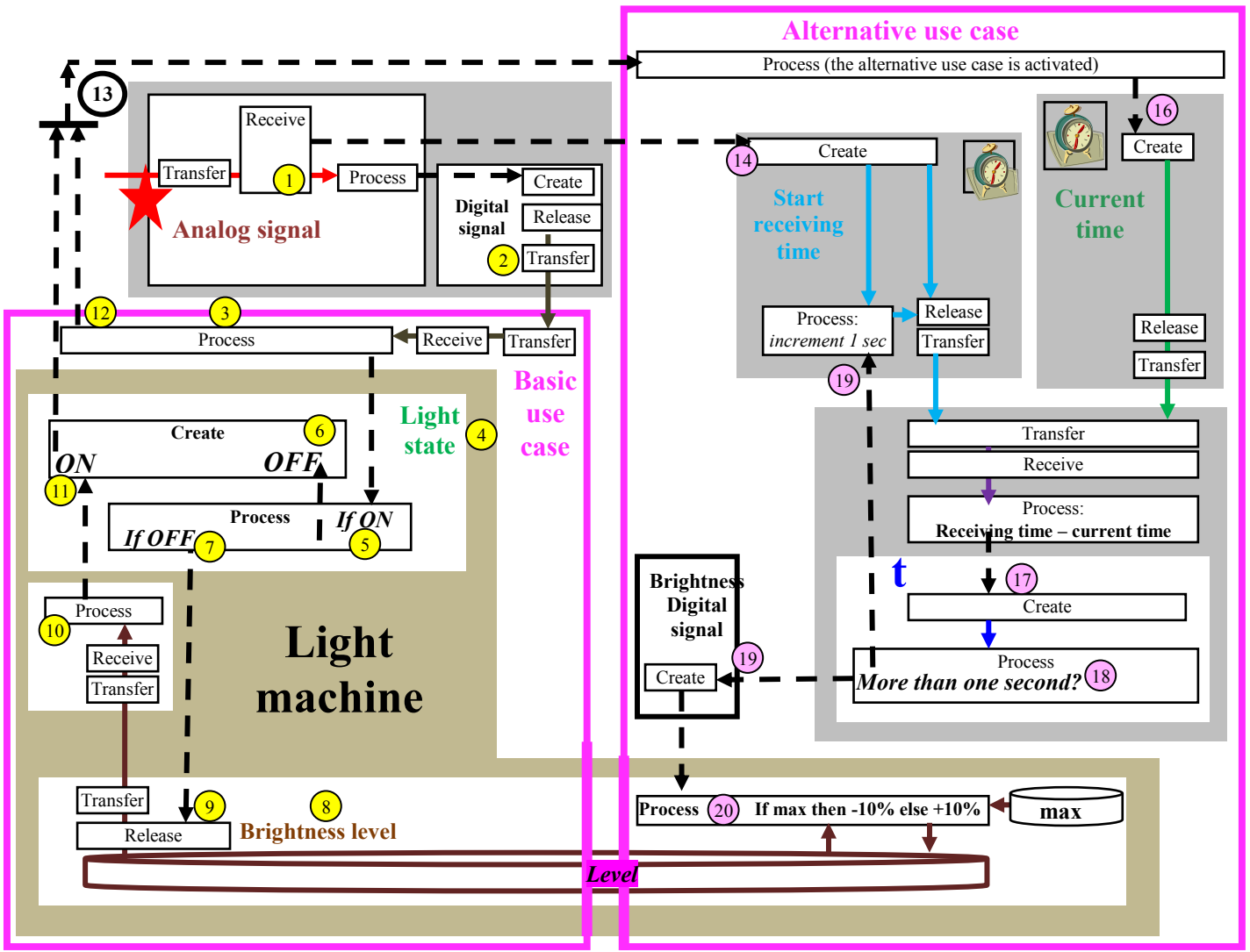

**Figure 15. TM model of the *Control Light* use case.**

2. The start of arrival time of the analog signal is registered (15).
3. Activating the alternative use case causes continuous retrieval of the current signal time (16).
4. t = Current time – starting time is calculated (17).
5. T is processed (19), and if t is greater than one second,
   (i) Starting time is incremented by one (19).
   (ii) A new brightness level is created (20). Note that we need here a flag (not shown) to indicate negative/positive change in the brightness level.
6. The process "The brightness level (8) is retrieved (9) and processed (10) to create an ON state (11)" mentioned in point 1 (iii) is repeated. The alternative use case is activated again. This process stops when t is not greater than one second (i.e., the resident is no longer pressing the button).

Fig. 16 identifies events in the *Control Light* use case as follows:

Event 1 ($E_1$): An analog signal reaches the lighting system, converted to digital signal and processed.
Event 2 ($E_2$): The current state is ON, and an OFF state is created.
Event 3 ($E_3$): The current state is OFF.
Event 4 ($E_4$): The current state is ON.
Event 5 ($E_5$): The brightness level is retrieved to turn on the light accordingly.
Event 6 ($E_6$): The process of recalculating the brightness is triggered because the input signal is still input.
Event 7 ($E_7$): Start time of the arrival of the signal is created.
Event 8 ($E_8$): Current time is also created.
Event 9 ($E_9$): t = Current time – starting time is calculated and the result is compared with one second.
Event 10 ($E_{10}$): Starting time is decremented by one.
Event 11 ($E_{11}$): Brightness level is calculated and stored.

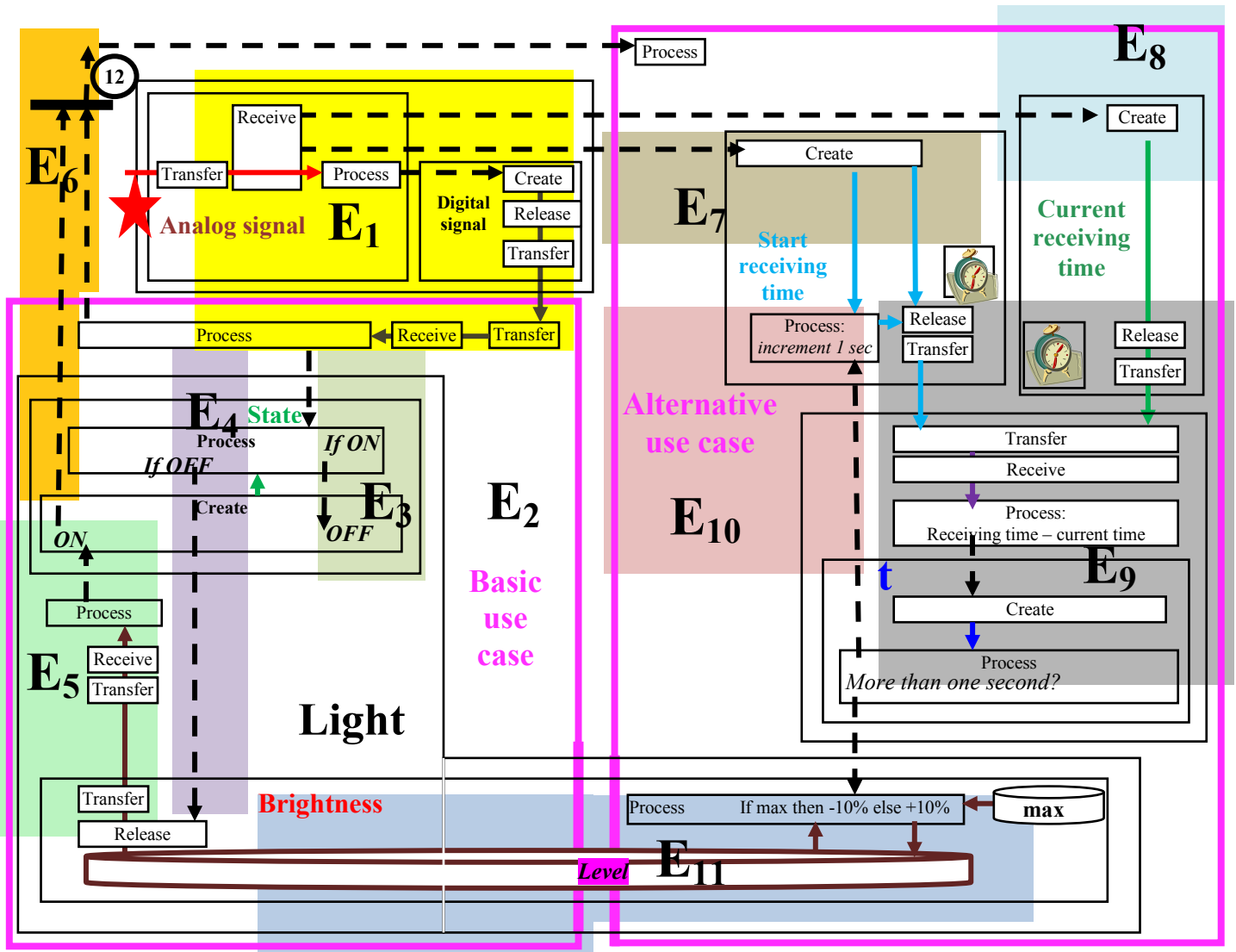

**Figure 16. TM model of the *Control Light* use case.**

Fig. 17 shows the chronology of events in the *Control Light* use case. Thus, use cases appear to be large events that have some special features to be "picked up" by the human mind while ignoring other structures of the same set of events. Fig. 18 shows the basic use case in this example as an event.

### V. USE CASES AS TRIGGERED MACHINES

Deshmukh [37] presented a valuable lab manual for visual modeling that includes use cases for the ATM system.

**System Startup Use Case**: The system initiates when the operator turns the operator switch to the "on" position. The operator will be asked to enter the amount of money currently in the cash dispenser, and a connection to the bank will be established. Then, the servicing of customers can begin [37].

**System Shutdown Use Case**: The system is shut down when the operator makes sure that no customer is using the machine and then turns the operator switch to the "off" position. The connection to the bank will be shut down. Then the operator is free to remove deposited envelopes, replenish cash and paper, etc. [37].

Fig. 19 shows the TM model of these use cases.
1. The operator action (1) triggers the switch to be either ON or OFF (2), and that puts the ATM in the ON or OFF state (3).
2. In the ON state (3), the startup begins (4).
   (a) The operator is asked to enter the amount of money currently in the cash dispenser (5 and 6), and he/she does that (7 and 8).
   (b) A connection is established with the bank (9).
   (c) The servicing of customers begins (10).
3. In the OFF state (3), the shutdown begins (11).
   (d) The connection to the bank is shut down (11).
   (e) Other things (e.g., deposited envelopes, cash, and paper) are removed or added (12).

Accordingly, we identify the following events in the two use cases, as shown in Fig. 20.

Event 1 ($E_1$): The operator action triggers the switch to be in the ON position, which puts the ATM in the state of ON; hence, the startup begins.

Event 2 ($E_2$): The operator is asked to enter the amount of money currently in the cash dispenser, and he/she does that.

Event 3 ($E_3$): A connection is established with the bank.

Event 4 ($E_4$): The servicing of customers begins.

Event 5 ($E_5$): The operator's action triggers the switch to be in the OFF position, and that puts the ATM in the state of OFF; hence, the shutdown begins.

Event 6 ($E_6$): The connection to the bank is shut down.

Event 7 ($E_7$): Other things (e.g., deposited envelopes, cash, and paper) are removed or added.

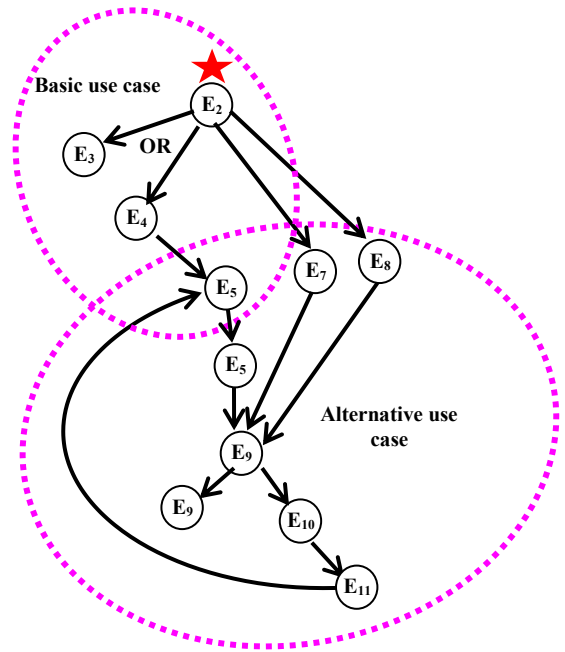

**Figure 17. Behavior of *Control Light* use case.**

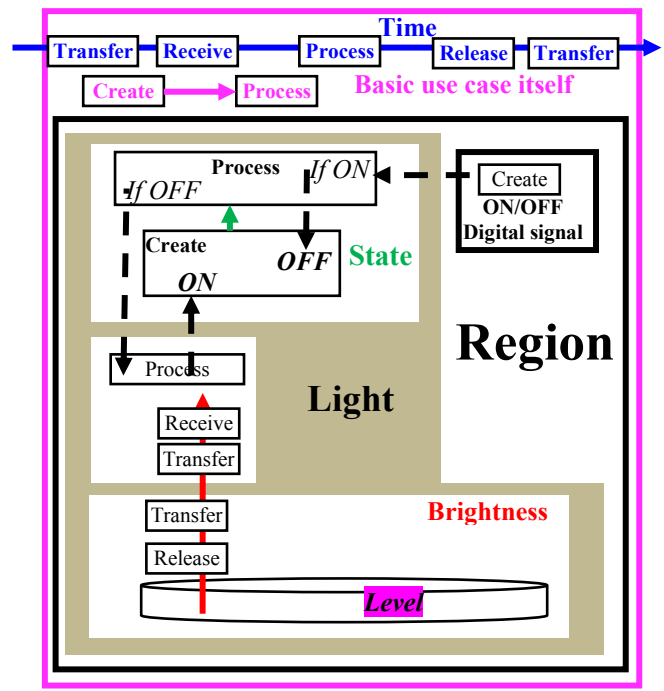

**Figure 18. Basic *Control Light* use case as a large event.**

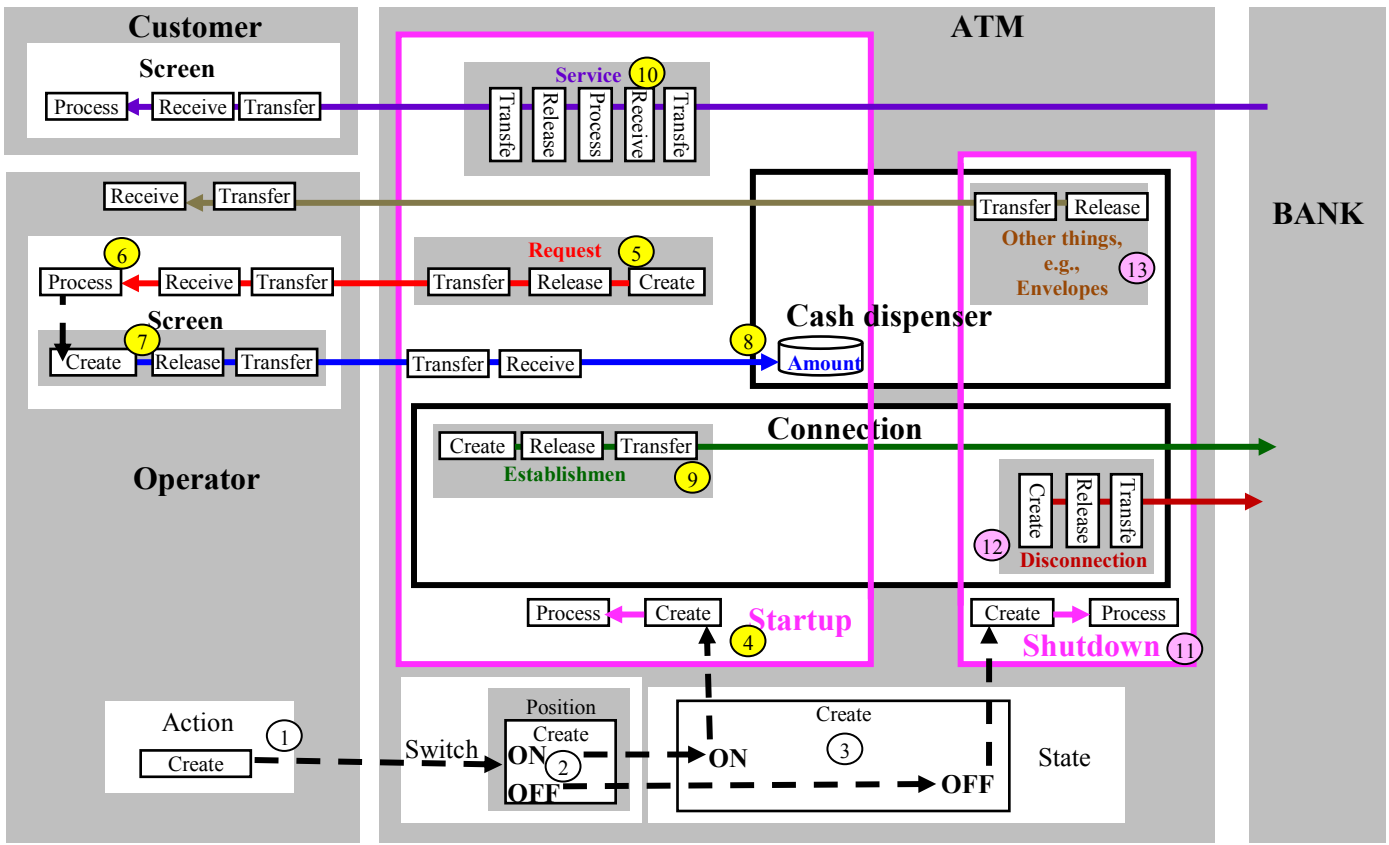

**Figure 19. TM model of the ATM use case.**

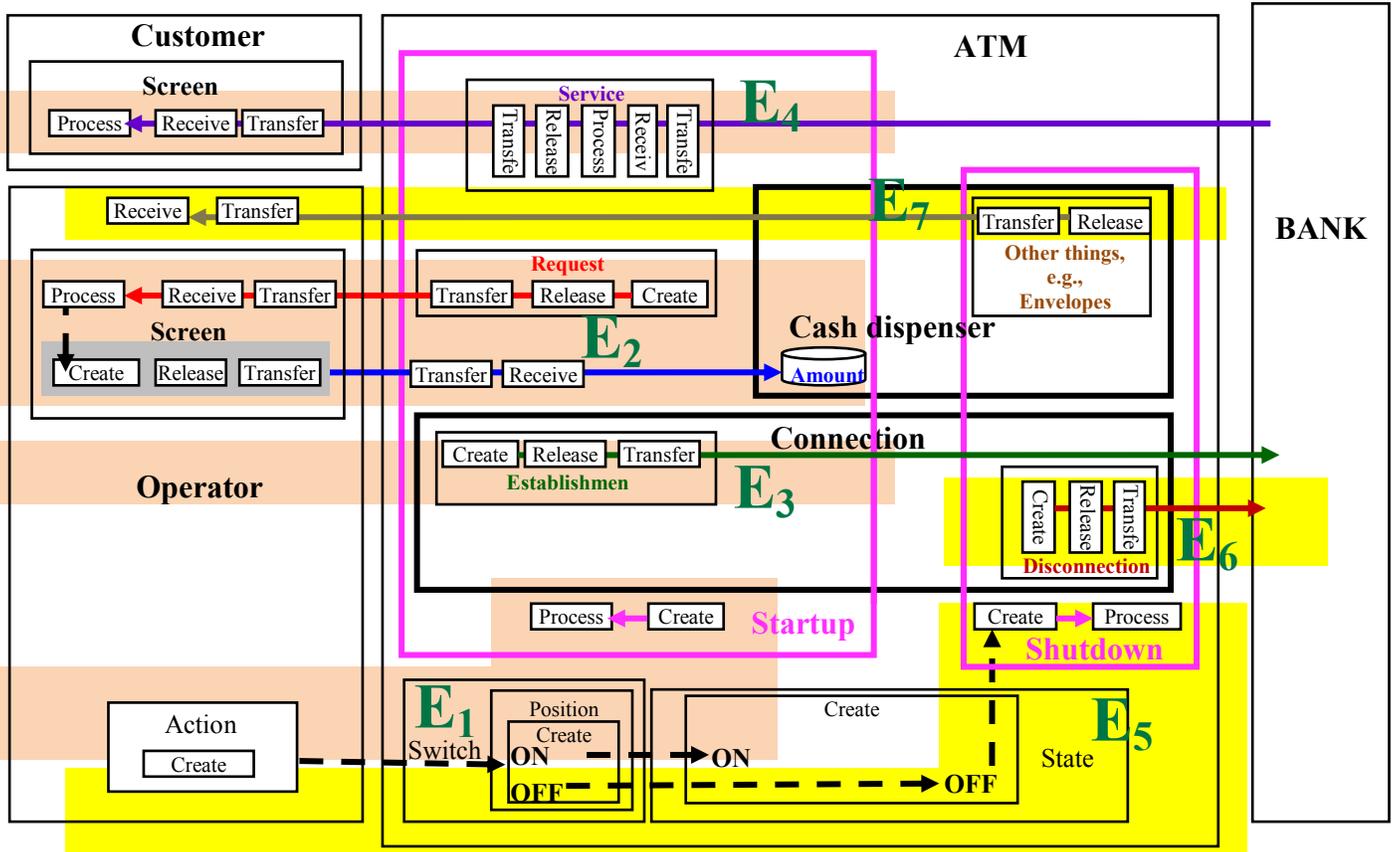

**Figure 20. TM model of the ATM use case.**

Fig. 21 shows the chronology of events in the two use cases.

Note that the so-called use case is a type of mega-triggering where a group of submachines are activated. The startup use case activates the submachines: Requesting a cash amount in the cash dispenser establishes a connection with the bank and begins the customer service. The shutdown use case activates the disconnection and removes the other things.

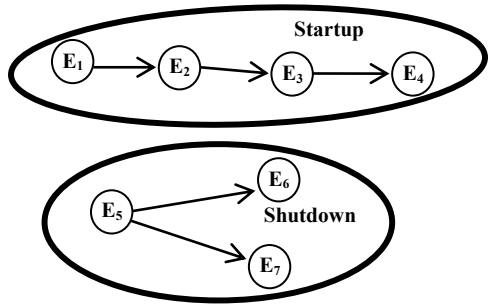

**Figure 21. The behavior of the ATM startup/shutdown.**

## VI. CONCLUSION

In this paper, we describe use cases in terms of a recently proposed model: the TM model. We remodeled selected use cases using the TM. This resulted in many insights related to the nature of use cases. The main result shows that use cases are big events wherein a group of submachines (processes) are activated. Events signify changes and have different levels of abstractions. The five generic processes in a TM denote basic changes in the TM model. Subsets of these changes form bigger events. For example, Release, Transfer, and Receive may form the larger event *sending something from one TM machine to another*. Accordingly, the events are structured from basic ones to larger events. Use cases are special large events that require human beings to select suitable semantic levels.